\documentclass[aps,prl,twocolumn]{revtex4}

\usepackage{bm}
\usepackage{color}
\usepackage{graphicx}
\usepackage{amsmath}

\begin{document}

\title{\Large{`Quantum weirdness' in exploitation by the international gravitational-wave observatory network}}

\author{Roman Schnabel$^{1}$} 
\affiliation{
$^{1}$ Institut f\"ur Laserphysik \& Zentrum f\"ur Optische Quantentechnologien, Universit\"at Hamburg, Luruper Chaussee 149, 22761 Hamburg, Germany
}

\email{roman.schnabel@physnet.uni-hamburg.de}

\maketitle 

\section*{Abstract}
{\bf{The detectors of the international gravitational-wave (GW) observatory network are currently taking data with sensitivities improved via squeezing the photon counting noise of the laser light used. Several GW candidate events, such as black-hole mergers, are already in the pipeline to be analyzed in detail. While the brand-new field of GW astronomy relies on squeezed light for reaching higher sensitivities, the physical understanding of such light, although being well-described by quantum theory, is still under discussion. 
Here, I present a description of why squeezed light, as now being exploited by GW observatories, constitutes rather remarkable physics. I consider the squeezed photon statistics and show its relation to the famous gedanken experiment formulated by Einstein, Podolsky, and Rosen in 1935. My description illuminates `quantum weirdness' in a clear way and might be the starting point for finding the physics of quantum correlations in general, which scientists have been seeking for decades. 
}}  

 \vspace{-3mm}
\section*{Introduction} \vspace{-3mm}
Gravitational-wave (GW) observatories are Michelson interferometers, which employ ultra-stable laser light that is reflected back and forth between 40\,kg-sized super-polished mirrors that are suspended in vacuum as pendulums. They pick up periodic relative arm-length changes of clearly below $10^{-21}$ at audio-band frequencies. On September 14, 2015, Advanced LIGO made the first ever observation of a GW signal, which was produced by two merging black holes at a distance of 1,3 billion light years (410 Mpc) \cite{GW150914}.

\begin{figure}[]
     \vspace{0mm} \hspace*{-3mm}
    \includegraphics[width=9.4cm]{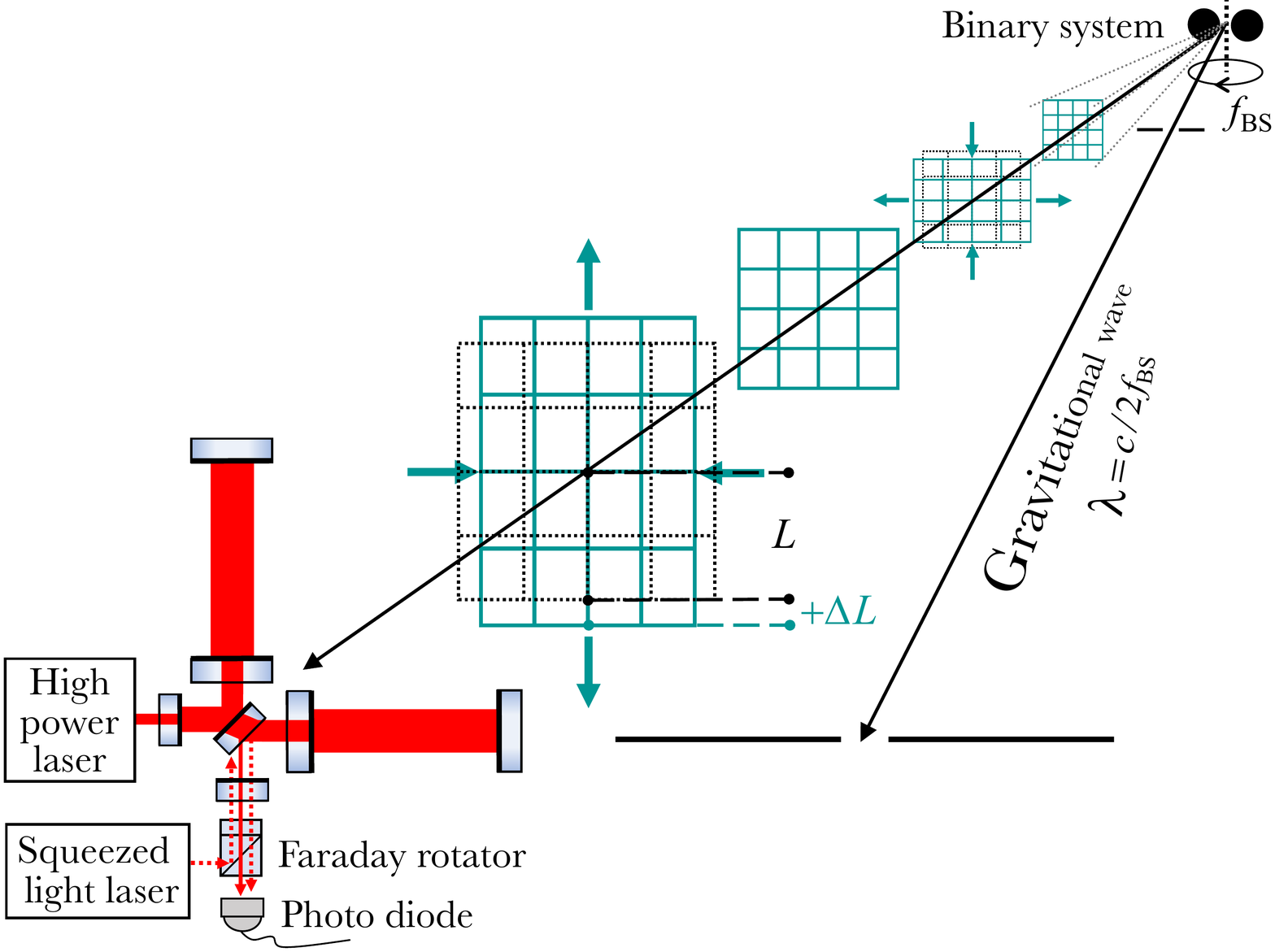}
    \vspace{-4mm}
    \caption{{\bf Gravitational-wave astronomy} -- Starting top right: Source of a gravitational wave in terms of a black-hole binary system with orbital frequency $f_{\rm BS}$, its effect on space-time geometry and its detection by a resonator- and squeezed-light enhanced GW observatory, such as those in the ongoing observational run.}
    \label{fig:1}
\end{figure}

One of the noise sources that limits the sensitivity of GW detectors is photon counting noise. It appears in the course of the photo-electric detection of the laser light by a photo diode in the interferometer's signal output port, see Fig.\,\ref{fig:1}. The possibility of squeezing the quantum uncertainty of the light's electric field in such a way that the photon counting statistics are smoothed was first theoretically found in 1976 by H. Yuen \cite{Yuen1976}. Five years later, squeezing was proposed for GW observatories \cite{Caves1981}. The technique proposed was eventually shown to be the optimum practical approach for interferometry with high precision on absolute scales \cite{Demkowicz-Dobrzanski2013}. All these aspects of squeezed states are fully described within the frame-work of quantum theory. For further reading on squeezed quantum states of light I refer to \cite{Walls1983,Breitenbach1997a,Schnabel2017}.

In 2010, Vahlbruch \emph{et al.} \cite{Vahlbruch2010} realised the first turn-key squeezed-light source with the aim to improve the sensitivity of a GW detector. One year later, this source was an integrated part of the GW detector GEO\,600 \cite{LSC2011}, and improved its sensitivity during GW searches in the so-called \emph{Astrowatch program} from 2011 to 2015 \cite{Grote2013}, when the LIGO and Virgo detectors were offline pursuing their respective advanced detector upgrade programs. The progress on squeezed-light sources \cite{Schnabel2010} led to the decision to go beyond the initially planned upgrade programs. In 2018, squeezed-light sources were regularly implemented in the Advanced LIGO and Advanced Virgo detectors. The advanced detectors' third concerted observational run started on April 1$^{\rm st}$, 2019. Since then, about one candidate event per week has been observed, as listed in the \emph{Gravitational-Wave Candidate Event Database} \cite{GraceDB}.

Here, I report on the historically unique situation that a new branch of astronomy depends on squeezed light for reaching higher sensitivities, while the comprehensive physical understanding of such light has still not been found. I present a clear description of why squeezed light indeed constitutes rather remarkable physics. My description is based on the Gaussian photon statistics of a single and bright squeezed beam of light. It directly incorporates the quantized energy transfer from the light to the photo-electric measurement device, which is not the case if electric-field quadrature amplitudes are considered, as it was done in Refs.\,\cite{Ou1992,Bowen2004}. The squeezed photon statistics in GW-observatory output-beams allow me to illuminate `quantum weirdness' in a surprisingly clear way.

\begin{figure}[]
     \vspace{-3mm}
    \includegraphics[width=9.1cm]{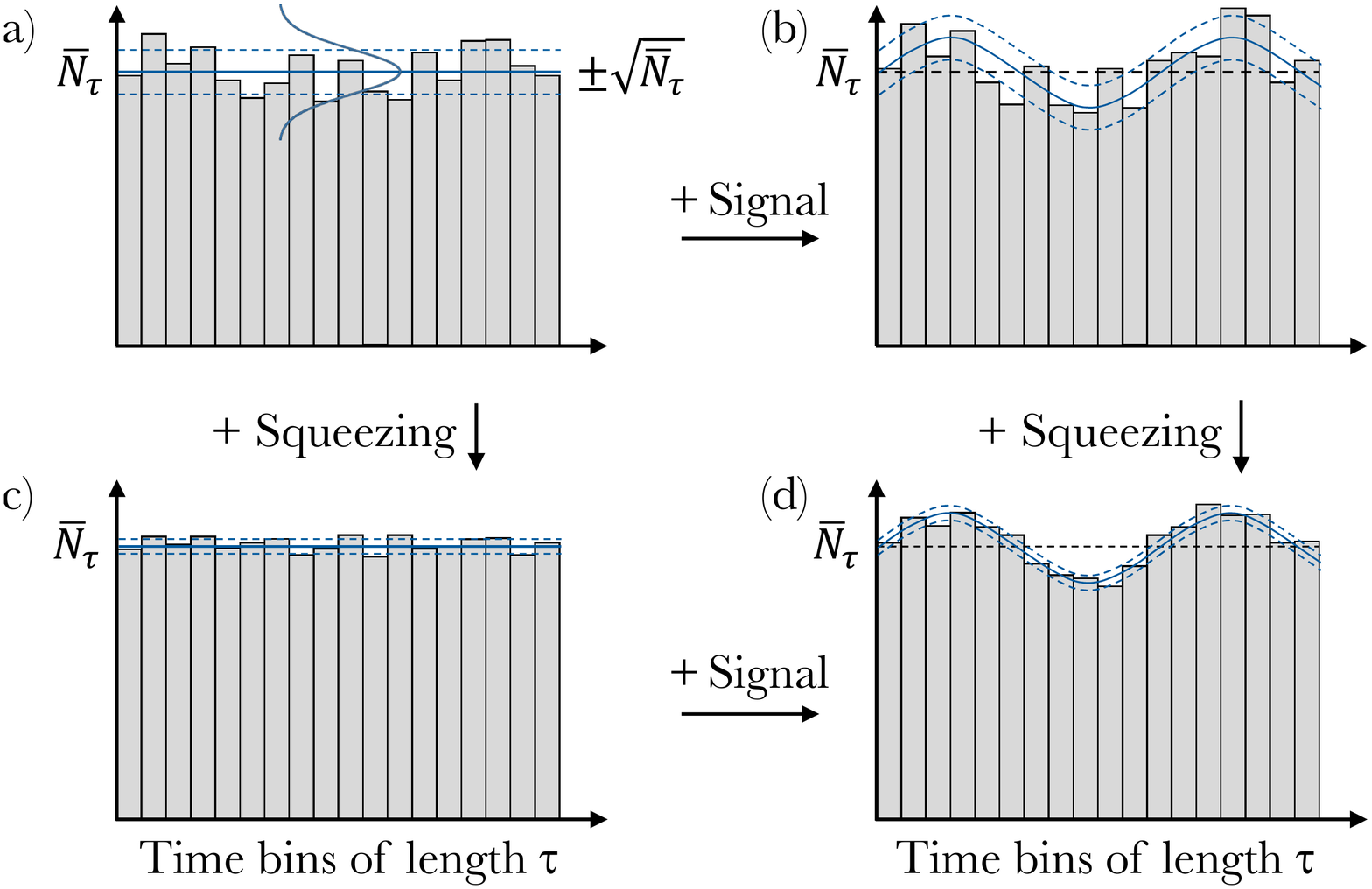}
    \vspace{-7mm}
    \caption{{\bf Illustration of photon counting noise} -- (a) For conventional laser light, the coherent quantum state provides the lowest level of photon counting noise. The fluctuation is solely due to the statistics of random, mutually independent photon detection events. For a large average number of $N$ photons per time bin, the standard deviation of the Gaussian distribution is $\sqrt{N}$. (b) Same photon counting noise together with a sinusoidal GW signal. (c) Reduced photon counting noise from laser light of the same power but in a squeezed state (here with a squeeze factor of about 10). (d) Similarly squeezed noise together with the same signal as in (b). The signal to noise ratio is improved. Quantum noise squeezing requires correlations in the photon detection events.}
    \label{fig:2}
\end{figure}

 \vspace{-4mm}
\section*{Squeezed photon statistics in GW observatories} \vspace{-3mm}
The observatories of the LIGO Scientific Collaboration (two Advanced LIGO detectors and GEO600) as well as the detector of the Virgo collaboration (Advanced Virgo) are Michelson interferometers using continuous-wave laser light at a wavelength of 1064\,nm. 
They are operated close to perfect destructive interference in the signal output port. Just a few tens of milliwatts of light with an ideally patternless Gaussian beam profile falls on a single photo diode, and the alternating expansion and shrinking of spacetime due to the GW are directly visible as a light-power change \cite{GW150914}. The photo diode produces an electric signal by converting ideally every photon into exactly one conduction electron. 
 
The key feature of squeezed light is its ability to improve the sensitivity of GW observatories without increasing the light power in its arms, thereby avoiding associated problems  \cite{Luck2000,Braginsky2001a,Evans2015,Vahlbruch2007,Accadia2010,Ottaway2012}.
Squeezing the photon statistics of the GW observatory output field requires an additional, completely new type of light source (Fig.\,\ref{fig:1}) \cite{Caves1981}. This source provides a squeezed vacuum field of almost zero light power, which is injected into the interferometers signal output port, is spatially overlapped with the conventional light field, and finally leaves the interferometer output port superimposed with the conventional field that carries the GW signal. The relative phase of the two fields needs to be stabilized such that the extrema of the conventional field superimpose the squeezed electric field uncertainties. The strength of the GW signal is not influenced.
Figure \ref{fig:2} shows Gaussian photon counting noise \emph{without} squeezing (a) and \emph{with} squeezing (c), both in the absence of a GW signal. For signal frequencies at which no other noise is relevant, such as from seismic or Brownian motion of the mirror surfaces, the signal to photon counting noise improves as shown in the transition from Fig.\,\ref{fig:2}\,(b) to (d).

 \vspace{-3mm}
\section*{The weirdness of squeezed photon statistics} \vspace{-3mm}
At first glance, measurements on systems with squeezed quantum uncertainties, as illustrated in Fig.\,\ref{fig:2}(c,d) or in diagrams showing distributions of electric field strengths of the light \cite{Walls1983,Breitenbach1997a,Schnabel2017}, do not seem to pose any fundamental question. In the following we carve out the remarkable (`weird') property of squeezed light on the basis of its (Gaussian) photon statistics. 

We first consider \emph{coherent} states, i.e.~a conventional intense continuous-wave laser beam, as it was used in every GW observatory before squeezed-light sources were added. (This setting is still an option if the output of the squeezed-light source in Fig.\ref{fig:1} is blocked).
For our discussion, the following property of the light in the interferometer's output port is of high relevance: it is extremely monochromatic. 
Monochromatic light is necessary for any GW observatory to allow for efficient coupling the light to the resonator system. 
Furthermore, frequency fluctuations would result in excess power fluctuations in the output, since the light's storage times in the arms cannot be perfectly identical for technical reasons.
For a quasi-monochromatic wave, the mathematics of Fourier transformation enforces a huge time spread of the wave packet, i.e.~it enforces a huge coherence time. The canonical example is a wave with an angular frequency spectrum of Gaussian shape and with standard deviation $\Delta \omega_a$, where $a$ denotes `amplitude'. In this case, the wave packet has a time spread with Gaussian envelope and standard deviation $\Delta t_a = 1/ \Delta \omega_a$. This equation sets the lower limit for the standard deviations of any wave. A wave packet at this limit is said to be a `Fourier-limited time-frequency mode'.
 
The mathematics of the Fourier transformation, together with the fact that interaction between a mode and a thermal bath is associated with energy exchange of no less than one photon of energy $E_{\rm Phot} = \hbar \omega$ \cite{Planck1900}, with $\hbar$ the reduced Planck constant, lead to the corresponding Heisenberg uncertainty relation. 
To show this, we square the absolute values of the wave packet's electric field as well as its Fourier transform considered in the example above. We get again a Gaussian envelope and a Gaussian, respectively, with new standard deviations $\Delta t = \Delta t_a / \sqrt{2}$ and $\Delta \omega = \Delta \omega_a /\sqrt{2}$. 
The two are energy distributions and, due to quantization, identical to photon number probability distributions, namely, versus photon detection time $t$ and photon angular frequency $\omega$. 

At this point, the following insight is essential. Since \emph{one} photon represents the interaction of a mode with a thermal bath, this photon must represent all properties of the mode. If the photon had the properties of another mode, then it would represent the interaction of the other mode with the thermal bath.
Standard deviations of the mode's energy distributions thus turn into uncertainties of a \emph{single} quantum and become `quantum uncertainties'. $t$ and $\omega$, as defined here, are operators, and the corresponding Heisenberg uncertainty relation reads $\Delta t \cdot \Delta (\hbar \omega) \geq \hbar/2$.    
(Alternatively, the same physics can be expressed as a longitudinal position/momentum uncertainty $\Delta z/c \cdot \Delta (c \hbar k_z) =  \Delta z \cdot \Delta p_z \geq \hbar/2$, where $\hbar k_z = p_z$ is the photon momentum along the optical axis of the laser beam.) 

The next important step for carving out the quantum weirdness of squeezed photon statistics is the fact that photon events in the observatory output are sampled into time bins of length $\tau$ being much shorter than the time uncertainty of the light $\Delta t$. Short time bins, say $\tau \lesssim 1$\,ms, are necessary to resolve GW frequencies of potentially up to kHz. Long coherence times, say $\Delta t \gtrsim 0.1$\,s, are required to keep the observatory noise low in the entire audio band.
In the previous paragraph we showed that the related Heisenberg uncertainty relation, as it is derived from Fourier mathematics and quantization, forbids the existence of photon arrival time information on short time-scale $\tau$ for light with a narrow-band spectrum of $1/(2\Delta t)$, if $\tau \ll \Delta t$. The photon detection events in short time bins must therefore be truly random, i.e.~they must show a Poissonian counting statistics. 
This is the result of this paragraph. 
The existence of true randomness, which is a key property of quantum physics, is nicely reproduced by combining Fourier mathematics and interaction quantization.

We now consider an interferometer that uses squeezed states of light. The first question to ask is whether the spectral width of the output light changes due to the superposed squeezed vacuum as shown in Fig.~\ref{fig:1}. The average number of photons added is given by $n = {\rm sinh}^2 r$ \cite{Walls1983}, where $r$ is the squeeze parameter \cite{Loudon1987}. For typical values of about ${\rm e}^{2r} \lesssim 10$, which corresponds to $\lesssim 10$\,dB \cite{Vahlbruch2008}, the added photon number is of the order of $1/$(sHz) \cite{Schnabel2017}. For a GW observatory with a squeezed signal-band from 10\,Hz to 10\,kHz this adds up to about 10$^4$ photons/s. In contrast, a few tens of milliwatts at 1064\,nm are equivalent to about $10^{17}$ photons/s. Thus the conventional quasi-monochromatic light clearly dominates the light's spectral width, and the answer to our question is `no'. This result is supported by the formal argument that the superposed squeezed vacuum does neither change the mode nor its properties but just its quantum state. The conclusion of this paragraph is the fact that squeezing the quasi-monochromatic light does not introduce the possibility to assign information about photon arrival times being more precise than $\Delta t$. 

The afore mentioned `weirdness' arises because the often-demonstrated squeezing of photon statistics, as illustrated in Fig.\,\ref{fig:2}, nevertheless prove the possibility of making rather precise predictions on photon numbers on short arrival time intervals $\tau \ll \Delta t$: 
When $N_{\tau,A}$ photons have been measured in short time bin $\tau_A$ around time $t_A$ (i.e.~on mode `$A$' ), where $t_A$ is defined with respect to the thermal bath of the measurement device, the value $N_{\tau,A}$ simultaneously is a relatively precise prediction of the photon numbers of subsequent time bins (i.e.~modes `$B$', `$C$' etc.). The prediction is obviously more precise than the best prediction in the reference case of the Poisson distribution of individually random photon arrivals.

The weird conflict between physically undefined photon arrival times and the possibility of predicting them finds its analogue in a sentence of the abstract in the seminal EPR paper \cite{Einstein1935}. It says `\emph{Consideration of the problem of making predictions concerning a system on the basis of measurements made on another system [...] one is [...] led to conclude that the description of reality as given by a wave function is not complete.} Indeed, 85 years ago, EPR would have perceived the observation of a squeezed photon counting statistics as a hint that quantum theory was not complete. 
It is probably reasonable to state that EPR overlooked the fact that Heisenberg's uncertainty relation, as it is based on the mathematics of the Fourier transformation, does not allow for a theory that assigns more precise information to individual quanta than quantum theory does. 
The experimental violation of Bell inequalities \cite{Aspect1981,Aspect1999,Giustina2013,Hensen2015} is a broadly accepted proof that quantum theory cannot be completed by information carried by so-called `local hidden variables'. 
Nevertheless, a comprehensive physical understanding neither exists for the violation of Bell inequalities nor for the squeezed photon statistics as employed by GW observatories.

 \vspace{-5mm}
\section*{Summary} \vspace{-3mm}
Squeezed light is successfully improving gravitational-wave astronomy. The mechanism is well-described by quantum theory, however, 
constitutes an example of `quantum weirdness'. The latter term has been used to describe a list of experiments, including gedanken experiments, that all lack a comprehensive physical understanding that a clear majority of physicists agrees on. I do not present the comprehensive physical understanding of squeezed photon statistics, but carve out -- on physical grounds -- the `quantum weirdness' in the observation of the time series with such a statistics. 
I consider the squeezed output-light of a GW observatory that is almost perfectly stable with respect to frequency and power over a time period $\Delta t$ (if no GW signal is present). Solely based on Fourier mathematics, I firstly show that the precise detection time of any photon \emph{within} $\Delta t$ cannot be physically defined with respect to the detectors's thermal bath, before the photon is actually detected. Secondly, I show that nevertheless rather precise predictions on future measurements of the photon number within time period $\tau \ll \Delta t$ are possible, based on a single such measurement.    
The weird conflict of indefiniteness before measurement and the possibility of predicting measurement outcomes on a system through a measurement on another system applies to all `quantum-weird' experiments. The clarity of the example of squeezed photon statistics might facilitate the endeavour for solving `quantum weirdness' in a way that will be broadly accepted in physics and beyond. 
 \vspace{-3mm}

\subsection*{Acknowledgements}\vspace{-3mm}
This work was funded by the Deutsche Forschungsgemeinschaft under Germany's Excellence Strategy -- EXC 2121 `Quantum Universe' -- 390833306
and by the European Research Council (ERC) Project MassQ (Grant No. 339897). This manuscript has the LIGO document number P1900282.

\end{document}